\documentclass[12pt]{article}

\usepackage{amsmath}
\usepackage{bm}
\def\beq{\begin{equation}}
\def\eeq{\end{equation}}
\def\beqa{\begin{eqnarray}}
\def\eeqa{\end{eqnarray}}
\def\bea*{\begin{eqnarray*}}
\def\eea*{\end{eqnarray*}}
\def\bc{\begin{center}}
\def\ec{\end{center}}
\def\bar{\overline}

\def\a{\alpha}
\def\b{\beta}

\def\G{\varGamma}

\def\D{\varDelta}
\def\e{\epsilon}

\def\l{\lambda}
\def\L{\varLambda}

\def\r{\rho}

\def\M{{\cal M}}

\def\O{{\cal O}}
\def\Z{{\cal Z}}

\bmdefine{\bw}{w}
\bmdefine{\bm}{m}
\bmdefine{\bz}{z}
\bmdefine{\bh}{h}
\bmdefine{\by}{y}

\def\PR#1#2#3{Phys. Rev.  {\bf #1}, #2 (#3)}
\def\PRL#1#2#3{Phys. Rev. Lett. {\bf #1}, #2 (#3)}
\def\PL#1#2#3{Phys. Lett. {\bf #1}, #2 (#3)}
\def\NP#1#2#3{Nucl. Phys. {\bf #1}, #2 (#3)}

\def\PTP#1#2#3{Prog. Theor. Phys. \textbf{#1}, #2 (#3)}

\begin{document}

\begin{center}
{\Large \bf Quark \textit{CP}-Phase and Froggatt-Nielsen Mechanism}

\vspace{15mm}

Chuichiro Hattori $^{\rm{1},}$
            \footnote{E-mail: yama\_houshi@ybb.ne.jp}, 
Masahisa Matsuda $^{\rm{2},}$\footnote{Permanent address : Aichi University of Education, Kariya 448-8542, Japan}
            \footnote{E-mail: mmatsuda@nara-edu.ac.jp; mmatsuda@auecc.aichi-edu.ac.jp},
Mamoru Matsunaga $^{\rm{3},\rm{4},}$
            \footnote{E-mail: matsuna@phen.mie-u.ac.jp; matsumamo@mediacat.ne.jp}, \\
and \\
Takeo Matsuoka $^{\rm{4},}$
            \footnote{E-mail: t-matsu@siren.ocn.ne.jp}
\end{center}

\vspace{5mm}

\begin{center}
\textit{
$^{\rm{1}}$Science Division, General Education, Aichi Institute of Technology, \\
     Toyota 470-0392, JAPAN \\
$^{\rm{2}}$Management Office, Nara University of Education, \\
     Nara 630-8528, JAPAN \\
$^{\rm{3}}$College of Liberal Arts and Sciences,  \\
     Mie University, Tsu 514-8507, JAPAN  \\
$^{\rm{4}}$Department of Physics Engineering, Mie University, \\
     Tsu 514-8507, JAPAN
 }
\end{center}

\vspace{10mm}

\begin{abstract}
On the basis of the Froggatt-Nielsen mechanism, 
we study quark flavor mixings in the  $SU(6) \times SU(2)_R$ model. 
The characteristic structure of the CKM matrix is attributed to the hierarchical 
effective Yukawa couplings due to the Froggatt-Nielsen mechanism and also 
to the state-mixings beyond the MSSM. 
We elucidate the detailed form of the CKM matrix elements and find 
interesting relations between the \textit{CP} violating phase and three mixing angles. 
Taking the existing data of three mixing angles, 
we estimate the quark \textit{CP}-phase at $\delta = (75 \pm 3)^{\circ}$. 
This result is in accord with observations. 
\end{abstract}

\newpage


\section{Introduction}

One of theoretically challenging issues is to understand characteristic features of 
quark mass patterns and the CKM-mixing matrix\cite{CKM}. 
It seems that the important key to this issue is the state-mixing 
between quarks and extra particles beyond the minimal supersymmetric standard 
model(MSSM). 
In fact, it was shown in the context of $SU(6) \times SU(2)_R$ string-inspired model, 
which contains massless particles beyond the MSSM, that we were able to explain 
characteristic patterns of the observed mass spectra and mixing matrices of 
quarks and leptons\cite{Matsu1,Matsu2,Matsu3,Matsu4,Matsu5}. 
In the model the Froggatt-Nielsen (F-N) mechanism\cite{F-N} plays an important role. 
It is noticeable that  doublet Higgs and color-triplet Higgs 
fields belong to different representations of $SU(6) \times SU(2)_R$. 
This situation is favorable to solve the triplet-doublet splitting problem. 
In addition, the longevity of the proton can be guaranteed under appropriate flavor 
symmetries\cite{Matsu1}.
In this paper we focus our attention on the detailed form of the CKM matrix elements 
in the above-mentioned model. 
It has been shown that in the model the hierarchical pattern of three mixing angles 
can be understood systematically\cite{Matsu3}. 
We shed light on relations between the \textit{CP} violating phase and three mixing angles 
in this paper.

In the present model, it is assumed that the hierarchical structure of fermion mass matrices 
is attributed to the F-N factors coming from the F-N mechanism. 
In the previous work\cite{HMM1}, we derived the typical relations among CKM matrix elements 
\beqa
\label{ckm1}
  |V_{cd}|  & = &  |V_{us}|,    \\
\label{ckm2}
  |V_{ts}|  & = &  |V_{cb}|,    \\
\label{ckm3}
  |V_{td}|  & = &  |V_{us} \, V_{cb}|. 
\eeqa
The CKM-matrix is defined as 
\begin{eqnarray}
   &  &  V_{\rm CKM} = \left(
                       \begin{array}{ccc}
                       \    V_{ud} \   &  \  V_{us}  \  &  \  V_{ub}    \\
                       \    V_{cd} \   &  \  V_{cs}  \  &  \  V_{cb}    \\
                       \    V_{td} \   &  \  V_{ts}  \  &  \  V_{tb}    
                       \end{array}
                        \right)   \nonumber  \\
             & = & \left(
                    \begin{array}{ccc}
                       c_{12} \, c_{13}    &    s_{12} \, c_{13}    &    s_{13} \, e^{-i\delta}    \\
                      -s_{12} \, c_{23} - c_{12} \, s_{23} \, s_{13} \, e^{i\delta}    
                  &    c_{12} \, c_{23} - s_{12} \, s_{23} \, s_{13} \, e^{i\delta}   &  s_{23} \, c_{13}    \\
                       s_{12} \, s_{23} - c_{12} \, c_{23} \, s_{13} \, e^{i\delta}    
                  &   -c_{12} \, s_{23} - s_{12} \, c_{23} \, s_{13} \, e^{i\delta}   &  c_{23} \, c_{13}   
                    \end{array}
                        \right)  \label{CKMPDG}
\end{eqnarray} 
in the standard representation of Particle Data Group(PDG)\cite{PDG}. 
Due to unitarity condition on $V_{\rm CKM}$, 
Eqs.(\ref{ckm1}) and (\ref{ckm2}) are equivalent to each other. 
From the independent relations (\ref{ckm2}) and (\ref{ckm3}), 
the \textit{CP}-phase $\delta$ is expressed in terms of the three mixing angles as 
\beqa
\label{1}
    \cos \, \delta & = & \frac{s_{23}^2 (s_{12}^2 - s_{13}^2) - s_{12}^2 \, c_{23}^2 \, s_{13}^2}
                        {2 \, s_{12} \, c_{12} \, s_{23} \, c_{23} \, s_{13}},    \\
\label{2}
    \cos \, \delta & = & s_{13} \, \frac{s_{12}^2 \, s_{23}^2 \, (1 + c_{13}^2) + c_{12}^2 \, c_{23}^2}
                        {2 \, s_{12} \, c_{12} \, s_{23} \, c_{23}}, 
\eeqa 
respectively. 
Here if we input the experimental values of 
$s_{12} = 0.22536 \equiv  \l, \ s_{23} \simeq \l^{2.1}$ and $s_{13} \simeq \l^{3.8}$\cite{PDG}, 
Eqs.(\ref{1}) and (\ref{2}) exhibit 
\beqa
  \cos \, \delta & \simeq & \frac{s_{12} \, s_{23}}{2 \, s_{13}} \simeq \l^{-0.7}/2,   \\
  \cos \, \delta & \simeq & \frac{s_{13}}{2 \, s_{12} \, s_{23}} \simeq \l^{0.7}/2, 
\eeqa
respectively. 
These results are incompatible with each other.

However, the relations (\ref{ckm2}) and (\ref{ckm3}) are derived in 
the leading approximation. 
So, we need to accomplish more accurate calculation in order to discuss 
the quark \textit{CP}-phase. 
For this reason, in this paper we carry out the analysis up to the next-to-leading 
approximation in the F-N scheme, 
which allows us to find more accurate relations among the CKM-matrix elements. 
For example, we obtain 
\bea*
    \left| V_{cd} \right|^2 \simeq \left| V_{us} \right|^2 
                            \times \left[ 1 - \left| V_{cb} \right|^2 \right] ,
\eea*
which yields an attractive relation between the \textit{CP}-phase and three mixing angles. 
Using these relations, we are able to estimate the quark \textit{CP}-phase 
without relying on a specific flavor symmetry.

This paper is organized as follows. 
In Sec. II we briefly explain Yukawa couplings in the $SU(6) \times SU(2)_R$ model 
together with the F-N mechanism. 
Solving the eigenvalue problem for the mass matrices of the up-type and down-type 
quark sectors, 
we derive the diagonalization matrices. 
In Sec. III the detailed form of the CKM matrix is presented and 
interesting equations among the CKM-matrix elements are found. 
In Sec.IV it is shown that these yield attractive interrelations 
between the quark \textit{CP}-phase and three mixing angles . 
Taking the existing data of three mixing angles, 
we estimate the quark \textit{CP}-phase at $\delta = (75 \pm 3)^{\circ}$, 
which is in accord with the current data of $\delta$. 
Section V is devoted to summary.

\vspace{5mm}

\section{Yukawa couplings and F-N mechanism}

Here we briefly summarize the parts of the model which are relevant to our analysis. 
For a more complete discussion, see Refs. \cite{Matsu1,Matsu2,Matsu3,Matsu4,Matsu5,HMM1}. 
In this model the unification gauge symmetry is assumed to be $SU(6) \times SU(2)_R$ 
at the underlying string scale $M_S$. 
The gauge group $G = SU(6) \times SU(2)_R$ is a subgroup of $E_6$. 
Within the framework of $E_6$ we assign three families and one vector-like 
multiplet to matter superfields, i.e., 
\beq
  3 \times {\bf 27}(\Phi_{1,2,3}) + 
        ({\bf 27}(\Phi_0)+\overline{\bf 27}({\bar \Phi})) .
\eeq
The superfields $\Phi$ are decomposed into two multiplets of $G$ as 
\beq
  \Phi({\bf 27})=\left\{
       \begin{array}{lll}
         \phi({\bf 15},{\bf 1}) & : & \quad \mbox{$\{Q,L,g,g^c,S\}$}, \\
         \psi(\overline{\bf 6},{\bf 2}) & : & \quad \mbox{$\{(U^c,D^c),(N^c,E^c),(H_u,H_d)\}$}, 
       \end{array}
       \right.
\eeq
where $g$, $g^c$ and $H_u$, $H_d$ represent colored Higgs and $SU(2)_L$-doublet 
Higgs superfields, respectively. 
Doublet Higgs and color-triplet Higgs fields belong to  different representations of 
$G$ and  this situation is favorable to solve the triplet-doublet splitting problem. 
The superfields $N^c$ and $S$ are {\it R}-handed neutrinos and $SO(10)$-singlets, respectively. 
Although $D^c$ and $g^c$ as well as $L$ and $H_d$ have the same quantum numbers under 
the standard model gauge group $G_{\rm SM} = SU(3)_c \times SU(2)_L \times U(1)_Y$, 
they belong to different irreducible representations of $G$. 
We assign odd (even) {\it R}-parity to superfields $\Phi_{1,2,3}$ ($\Phi_0$ and $\bar{\Phi}$). 
Since ordinary Higgs doublets have even {\it R}-parity, they are contained in $\Phi_0$. 
It is assumed that {\it R}-parity remains unbroken down to the electroweak scale.

The gauge symmetry $G$ gets spontaneously broken in two steps at the scales 
$\langle S_0\rangle=\langle \bar{S} \rangle$ and 
$\langle N_0^c\rangle=\langle \bar{N^c} \rangle$ to $G_{\rm SM}$ as 
\[
   G = SU(6) \times SU(2)_R 
     \buildrel \langle S_0 \rangle \over \longrightarrow 
             SU(4)_{\rm PS} \times SU(2)_L \times SU(2)_R  
     \buildrel \langle N^c_0 \rangle \over \longrightarrow 
     G_{\rm SM}, 
\]
where $SU(4)_{\rm PS}$ represents the Pati-Salam $SU(4)$ \cite{Pati}. 
The $D$-flatness conditions require $\langle S_0\rangle=\langle {\overline S} \rangle$ 
and $\langle N_0^c\rangle=\langle {\overline N^c} \rangle$ at each step of 
the symmetry breakings. 
Hereafter it is supposed that the symmetry breaking scales are 
$\langle S_0 \rangle = 10^{17 - 18}$GeV and $\langle N^c_0 \rangle = 10^{15 - 17}$GeV. 
Under the $SU(4)_{\rm PS} \times SU(2)_L \times SU(2)_R$ the chiral superfields 
$\phi({\bf 15},{\bf 1})$ and $\psi(\overline{\bf 6},{\bf 2})$ are decomposed as 
\begin{eqnarray*}
   ({\bf 15},{\bf 1})         &=& {\bf (4,2,1)} + {\bf (6,1,1)} + {\bf (1,1,1)}, \\
   (\overline{\bf 6},{\bf 2}) &=& {\bf (\overline{4},1,2)} + {\bf (1,2,2)}. 
\end{eqnarray*}
{}From the viewpoint of the string unification theory, it is probable 
that the hierarchical structure of Yukawa couplings is attributed to 
some kind of flavor symmetries at the string scale $M_S$. 
If the flavor symmetry contains Abelian groups, 
the F-N mechanism works for the interactions among quarks, leptons and Higgs 
fields. 
The superpotential at the string scale is governed by the flavor symmetry as well 
as the gauge symmetry $G$. 
Aside from the flavor symmetry, we have two types of gauge invariant trilinear 
combinations 
\beqa
    (\phi ({\bf 15},{\bf 1}))^3 & = & QQg + Qg^cL + g^cgS, \nonumber \\
    \phi ({\bf 15},{\bf 1})(\psi (\overline{\bf 6},{\bf 2}))^2 & 
            = & QH_dD^c + QH_uU^c + LH_dE^c  + LH_uN^c \\
             {}& & \qquad   + SH_uH_d + gN^cD^c + gE^cU^c + g^cU^cD^c.  \nonumber 
\eeqa
They must be multiplied by additional $G$-invariant factors suppressed by powers of 
$1/M_S$ to form flavor symmetric terms. 
Namely, the couplings arise from the nonrenormalizable terms controlled by 
the flavor symmetry \cite{Matsu1,Matsu6,Matsu7}.

We first consider the effective Yukawa couplings of up-type quark sector, 
which are given by 
\beqa
  W_U  = \sum_{i,j=1}^{3} \, \M_{ij} Q_i U^c_j H_{u0}. 
\eeqa
Due to the F-N mechanism, the dimensionless matrix $\M$ takes the form 
\beq
             \M = f_M \, \G_1 M \G_2. 
\eeq
Our basic assumption is that the hierarchical structure of all $3 \times 3$ mass matrices 
is attributed to the F-N factors $\G_1$ and/or $\G_2$. 
Hence, hierarchy of $\M_{ij}$ stems only from $\G_1$ and $\G_2$, and the dimensionless 
matrix $M$ contains no hierarchical structure. 
Here we put a factor $f_M$ in order to set $\det M = 1$. 
It means that all the elements of $M$ are $\O(1)$. 
The F-N factors $\G_1$ and $\G_2$ are described as 
\beq
    \G_1 = {\rm diag}( x^{\a_1}, \ x^{\a_2}, \ 1), \qquad 
             \G_2 = {\rm diag}( x^{\b_1}, \ x^{\b_2}, \ 1) 
\eeq
with the hierarchy $x^{\a_1} \ll x^{\a_2} \ll 1$ and $x^{\b_1} \ll x^{\b_2} \ll 1$. 
To be specific, we take the F-N factors like $x^{\a_1} \sim \l^3$,\, 
$x^{\a_2}\simeq \ x^{\b_2} \sim \l^2$ and $ x^{\b_1} \sim \l^{4 \-- 5}$ 
consonant to the experimental data.

The mass matrix $\M$ is diagonalized via biunitary transformation as 
\beq
    {\cal V}_u^{-1} \M \,{\cal U}_u = \L_u, \qquad 
                   v_{u0} \L_u = {\rm diag}(m_u, \ m_c, \ m_t) 
\eeq
with $v_{u0} = \langle H_{u0} \rangle$. 
According to the standard procedure for diagonalizing $\M \M^{\dag}$, we obtain mass eigenvalues 
\beqa
   (m_u, \ m_c, \ m_t) \, \simeq \, |v_{u0} \, f_M| \times 
             \left( \frac{1}{|\bar{m}_{11}|} \, x^{\a_1 + \b_1}, \ 
                     \frac{|\bar{m}_{11}|}{|m_{33}|} \, x^{\a_2 + \b_2}, \  |m_{33}| \right), 
\eeqa
where $m_{ij} = (M)_{ij}$, $\bar{m}_{ij} = (M^{-1})_{ji}^* = \Delta(M)_{ij}^*$. 
The diagonalization matrix ${\cal V}_u$ is described in terms of eigenvectors 
$\bw_i^{(u)}$ of $\M \M^{\dag}$ as 
\beqa
   {\cal V}_u = ( \bw_1^{(u)}, \ \bw_2^{(u)}, \ \bw_3^{(u)}), 
\eeqa
where $\bw_i^{(u)}$ are expressed as 
\beqa
      \bw_1^{(u)} = N_1^{(u)} \left(
              \begin{array}{c}
                     1      \\
                 u_1^{(u)}  \\
                 v_1^{(u)}  
              \end{array}
              \right),  \quad 
     \bw_2^{(u)} = N_2^{(u)} \left(
              \begin{array}{c}
                 u_2^{(u)}  \\
                     1      \\
                 v_2^{(u)}  
              \end{array}
              \right),  \quad 
     \bw_3^{(u)} = N_3^{(u)} \left(
              \begin{array}{c}
                 u_3^{(u)}  \\
                 v_3^{(u)}  \\
                     1      
              \end{array}
              \right). 
\eeqa
Here $N_i^{(u)}$ are normalization factors. 
The phase factors are so chosen that the diagonal elements of ${\cal V}_u$ are real. 
Explicit forms of $u_i^{(u)}$ and $v_i^{(u)}$ $(i = 1,  2,  3)$ are 
\begin{equation}
 \begin{split}
   u_1^{(u)} & =  x^{\a_1 - \a_2} \ \left[ \displaystyle\frac{\bar{m}_{21}}{\bar{m}_{11}} + 
                                           \O(x^{2(\b_1 - \b_2)}) \right],     \\
   v_1^{(u)} & =  x^{\a_1} \ \left[ \displaystyle\frac{\bar{m}_{31}}{\bar{m}_{11}} + 
                                           \O(x^{2(\b_1 - \b_2)}) \right],       \\
   u_2^{(u)} & =  - x^{\a_1 - \a_2} \ \left[ \displaystyle\frac{\bar{m}_{21}^*}{\bar{m}_{11}^*} + 
                                           \O(x^{2\a_2}, \ x^{2(\b_1 - \b_2)}) \right], \\
   v_2^{(u)} & =  - x^{\a_2} \ \left[ \displaystyle\frac{m_{23}^*}{m_{33}^*} - 
         x^{2(\a_1 - \a_2)} \, \displaystyle\frac{m_{13}^* \, \bar{m}_{21}^*}{m_{33}^* \, \bar{m}_{11}^*} 
         + \O(x^{2\b_2}) \right],        \\
   u_3^{(u)} & =  x^{\a_1} \ \left[ \displaystyle\frac{m_{13}}{m_{33}} 
                                      + \O(x^{2\b_2}) \right],       \\
   v_3^{(u)} & =   x^{\a_2} \ \left[ \displaystyle\frac{m_{23}}{m_{33}} 
                                             + \O(x^{2\b_2}) \right], 
 \end{split}
\end{equation}
where $x^{\a_1 - \a_2} \sim \l$, $x^{\a_1} \sim \l^3$ and $x^{\a_2} \sim \l^2$. 

Note that $x^{2(\b_1 - \b_2)}, \ x^{2\a_2}$ and $x^{2\b_2}$ are $\O(\l^4)$ or less than $\O(\l^4)$. 
The normalization factors are given by 
\begin{eqnarray}
   N_1^{(u)} & = &  1 - \, x^{2(\a_1 - \a_2)} \, \frac{|\bar{m}_{21}|^2}{2 |\bar{m}_{11}|^2} 
                                + \, \O(x^{4(\a_1 - \a_2)}),    \nonumber   \\
   N_2^{(u)} & = &  1 - \, x^{2(\a_1 - \a_2)} \, \frac{|\bar{m}_{21}|^2}{2 |\bar{m}_{11}|^2} 
                           + \, \O(x^{4(\a_1 - \a_2)}, \  x^{2\a_2}),    \\
   N_3^{(u)} & = &  1 + \, \O(x^{2\a_2}),   \nonumber 
\end{eqnarray}
and we have the relation $N_1^{(u)}N_3^{(u)}=N_2{(u)}(1+\O(\l^6))$.

We next proceed to study the effective Yukawa couplings of down-type quark sector, 
which are of the form 
\beqa
    W_D = \sum_{i,j=1}^{3} \, \left[ \Z_{ij} \ g_i g^c_j S_0 + 
          \M_{ij} \left( g_i D^c_j N^c_0 + Q_i D^c_j H_{d0} \right) \right], 
\eeqa
where ${\cal Z} = f_Z \, \G_1 Z \G_1$ and $\det Z = 1$. 
It is assumed that there is no hierarchical structure in $Z$. 
The mass matrix of down-quark sector is given by the $6 \times 6$ matrix 
\beqa
   \begin{array}{r@{}l} 
       \vphantom{\bigg(}   &  \begin{array}{ccc} 
            \quad  \ g^c   &  \quad \  D^c  &  
   \end{array}  \\ 
   \widehat{\M}_d = 
      \begin{array}{l} 
        g   \\  D  \\ 
      \end{array} 
       & 
   \left( 
      \begin{array}{cc} 
          \r_S \Z    &    \r_N \M   \\
             0       &    \r_d \M   
      \end{array} 
\right), 
\end{array} 
\eeqa
where $\r_S = \langle S_0 \rangle /M_S,  \r_N = \langle N^c_0 \rangle /M_S$ 
and $ \r_d = \langle H_{d0} \rangle /M_S = v_{d0} /M_S$. 
It is noticeable that $D^c$-$g^c$ mixings occur in down-type quark sector. 
Diagonalization is accomplished via biunitary transformation as 
\beqa
    \widehat{\cal V}_d^{-1} \widehat{\M}_d \, \widehat{\cal U}_d 
                   = {\rm diag}(\L_d^{(0)}, \ \e_d \, \L_d^{(2)}), 
\eeqa
where $\e_d = \r_d / \r_N = v_{d0} / \langle N^c_0 \rangle = \O(10^{-15})$. 
$\L_d^{(0)}$ means the heavy modes with the GUT scale masses. 
To solve the eigenvalue problem, we deal with $\widehat{\M}_d \widehat{\M}_d^{\dag}$, 
which are expressed as 
\beqa
   \widehat{\M}_d \widehat{\M}_d^{\dag} =  \left(
  \begin{array}{cc}
     A_d + B_d   &   \e_d^* B_d   \\
     \e_d B_d    &  |\e_d|^2 B_d  
  \end{array}
  \right) 
  \label{MMdag}
\eeqa
with the notation $A_d = |\r_S|^2 \, \Z \Z^{\dag}$ and $B_d = |\r_N|^2 \, \M \M^{\dag}$. 
Within $O({\epsilon_d}^2)$ mass eigenvalues $\L_d^{(2)}$ are given as 
\beqa
    ( \L_d^{(2)})^2 = {\cal V}_d^{-1} (A_d^{-1} + B_d^{-1})^{-1} {\cal V}_d 
\eeqa
and 
\beqa
     M_S \, |\e_d \, \r_N| \, \L_d^{(2)} = {\rm diag}(m_d, \ m_s, \ m_b), 
\eeqa
where $ {\cal V}_d $ is unitary  within $O(\epsilon_d)$ as seen in Eq.~\eqref{MMdag}.
It turns out that down-type quark masses are 
\beqa
   (m_d, \ m_s, \ m_b) \, \simeq \, |v_{d0} \, f_M| \times 
          \left( \frac{1}{\sqrt{l_{11}}} \, x^{\a_1 + \b_1}, \ 
               \sqrt{\frac{l_{11}}{g}} \, x^{\a_2 + \b_1}, \ 
                   \sqrt{\frac{g}{h}}  x^{\b_1} \right), 
\eeqa
where 
\beqa
    l_{ij}   & = & \xi_d^2 \, \bar{z}_{i1} \,\bar{z}_{j1}^* + \bar{m}_{i1} \,\bar{m}_{j1}^*, \nonumber \\
    g        & = & \xi_d^2 \, |D_3^{11}|^2,                                                \nonumber \\
    h        & = & \xi_d^4 \, x^{2(\a_1 - \a_2)} \, |({\bz}_3 \cdot \bar{\bm}_1^*)|^2 \, + \, 
                      \xi_d^2 \, x^{2(\b_1 - \b_2)} \, |({\bm}_3 \cdot \bar{\bz}_1^*)|^2,  \nonumber \\
    &  &  \xi_d^2 \ = \ \left| \frac{\r_N f_M}{\r_S f_Z} \right|^2 \, x^{2(\b_1 - \a_1)},  \qquad 
             D_k^{ij} = (\bar{\bz}_i \times \bar{\bm}_j)_k.                            \nonumber 
\eeqa
Here we use the notations $z_{ij} = (Z)_{ij}$, $\bar{z}_{ij} = \D(Z)_{ij}^*$ and 
\beqa
    &  &  {\bm}_i = ( m_{1i}, \ m_{2i}, \ m_{3i})^T, \qquad \ 
              \bar{\bm}_i = ( \bar{m}_{1i}, \ \bar{m}_{2i}, \ \bar{m}_{3i})^T, 
                                                                        \nonumber  \\
    &  &  {\bz}_i = ( z_{1i}, \ z_{2i}, \ z_{3i})^T, \qquad \qquad \ 
              \bar{\bz}_i = ( \bar{z}_{1i}, \ \bar{z}_{2i}, \ \bar{z}_{3i})^T.   \nonumber  
\eeqa
The diagonalization matrix ${\cal V}_d$ is expressed as 
\beqa
   {\cal V}_d = ( \bw_1^{(d)}, \ \bw_2^{(d)}, \ \bw_3^{(d)}) 
\eeqa
with 
\beqa
     \bw_1^{(d)} = N_1^{(d)} \left(
              \begin{array}{c}
                     1      \\
                 u_1^{(d)}  \\
                 v_1^{(d)}  
              \end{array}
              \right),  \quad 
     \bw_2^{(d)} = N_2^{(d)} \left(
              \begin{array}{c}
                 u_2^{(d)}  \\
                     1      \\
                 v_2^{(d)}  
              \end{array}
              \right),  \quad 
     \bw_3^{(d)} = N_3^{(d)} \left(
              \begin{array}{c}
                 u_3^{(d)}  \\
                 v_3^{(d)}  \\
                     1      
              \end{array}
              \right). 
\eeqa
Here the phase factors are so taken that the diagonal elements of ${\cal V}_d$ are real. 
Each element of ${\cal V}_d$ is of the form 
\begin{equation}
 \begin{split}
   u_1^{(d)} & =  x^{\a_1 - \a_2} \ \left[ \displaystyle\frac{l_{21}}{l_{11}} + 
                           x^{2(\a_1 - \a_2)} \, \frac{\xi_d^2}{(l_{11})^2} 
                       \Bigl( \bar{z}_{12}^* \, n_{13}^* 
                          + \frac{l_{21}}{l_{11}} \, | D_3^{11} |^2 \Bigr) + \O(\l^4) \right],   \\
   v_1^{(d)} & =  x^{\a_1} \ \left[ \displaystyle\frac{l_{31}}{l_{11}} - 
                          x^{2(\a_1 - \a_2)} \, \frac{\xi_d^2}{(l_{11})^2} 
                       \Bigl( \bar{z}_{12}^* \, n_{12}^* 
                   + \frac{l_{21}}{l_{11}} \, D_2^{11} \, D_3^{11*} \Bigr) + \O(\l^4) \right],   \\
   u_2^{(d)} & =  - \ x^{\a_1 - \a_2} \ \left[ \displaystyle\frac{l_{12}}{l_{11}} + 
                         x^{2(\a_1 - \a_2)} \, \frac{\xi_d^2}{(l_{11})^2} 
                       \Bigl( \bar{z}_{12} \, n_{13} 
                          + \frac{l_{12}}{l_{11}} \, | D_3^{11} |^2 \Bigr) + \O(\l^4) \right],         \\
   v_2^{(d)} & =  - \ x^{\a_2} \ \left[ \displaystyle\frac{ D_2^{11}}{ D_3^{11}} 
                                + x^{2(\a_1 - \a_2)} 
                       \Bigl( f_{\a}^* \, n_{13} 
                    - \frac{l_{12} \, D_1^{11}}{l_{11} \, D_3^{11}} \Bigr) + \O(\l^4) \right],    \\
   u_3^{(d)} & =  x^{\a_1} \ \left[ \displaystyle\frac{D_1^{11*}}{D_3^{11*}} 
                          - x^{2(\a_1 - \a_2)} f_{\a} \, n_{23}^* + \O(\l^4)  \right],     \\
   v_3^{(d)} & =  x^{\a_2} \ \left[ \displaystyle\frac{D_2^{11*}}{D_3^{11*}} 
                          + x^{2(\a_1 - \a_2)} f_{\a} \, n_{13}^* + \O(\l^4)  \right],  
 \end{split}
\end{equation}
where 
\[
    n_{ij} =  \xi_d^2 \, \bar{z}_{i1} \, z_{j3} - \bar{m}_{i1} \, D_j^{21*},  \qquad 
    f_{\a} = \displaystyle\frac{(\bz_3 \cdot \bar{\bm}_1^*)}{D_3^{11*} \, | D_3^{11} |^2}. 
\]
The normalization factors become 
\begin{eqnarray}
   N_1^{(d)} & = &  1 - x^{2(\a_1 - \a_2)} \, \frac{|l_{21}|^2}{2(l_{11})^2} 
                                                  + \, \O(x^{4(\a_1 - \a_2)}),    \nonumber   \\
   N_2^{(d)} & = &  1 - x^{2(\a_1 - \a_2)} \, \frac{|l_{21}|^2}{2(l_{11})^2} 
                                                  + \, \O(x^{4(\a_1 - \a_2)}, \ x^{2\a_2}),   \\
   N_3^{(d)} & = &  1 + \O(x^{2\a_2})    \nonumber 
\end{eqnarray}
with
\beqa
    N_1^{(d)}  N_3^{(d)} = N_2^{(d)} ( 1 + \O(\l^6) ). 
\eeqa

\vspace{5mm}

\section{The CKM matrix}

In the present framework the CKM matrix is given by 
\beqa
V_{\rm CKM}={\cal V}_u^{-1} \,{\cal V}_d = {\cal V}_u^{\dag} \,{\cal V}_d
\eeqa 
and each element of $V_{\rm CKM}$ becomes 
\beqa
(V_{\rm CKM})_{ij} = {\bw}_i^{(u)*} \cdot {\bw}_j^{(d)}. 
\eeqa
Using approximate analytic expressions of ${\bw}^{(u)}$ and ${\bw}^{(d)}$ given 
in the preceding section, 
we are in a position to exhibit each element of $V_{\rm CKM}$ explicitly. 
Thus $V_{us}$ and $V_{cd}$ are expressed as 
\begin{eqnarray}
  V_{us}  & = &  x^{\a_1 - \a_2} \, N_1^{(u)} N_2^{(d)} \, 
                   \left[ \frac{\xi_d^2 \, \bar{z}_{11} \, D_3^{11*}}{\bar{m}_{11}^* \, l_{11}} \ 
                      - \ x^{2(\a_1 - \a_2)} \, \frac{\xi_d^2}{(l_{11})^2} \times \right.  \nonumber  \\
          &  &  \qquad  \left. \times \Bigl( \bar{z}_{12} \, n_{13} 
                                + \ \frac{l_{12}}{l_{11}} |D_3^{11}|^2 \Bigr) + \O(\l^4)  \right],    \\  
  V_{cd}  & = &  - \ x^{\a_1 - \a_2} \, N_2^{(u)} N_1^{(d)} \, 
                   \left[ \frac{\xi_d^2 \, \bar{z}_{11}^* \, D_3^{11}}{\bar{m}_{11} \, l_{11}} \ 
                        - \ x^{2(\a_1 - \a_2)} \frac{\xi_d^2}{(l_{11})^2} \times  \right.  \nonumber \\
          &  &  \qquad  \left. \times \Bigl( \bar{z}_{12}^* \, n_{13}^* 
                                + \ \frac{l_{21}}{l_{11}} |D_3^{11}|^2 \Bigr) + \O(\l^4)  \right], 
\end{eqnarray}
where $x^{\a_1 - \a_2} \sim \l$. 
The relation of $V_{cd} = - (V_{us})^* \times (1 + \O(\l^4))$ is hold 
in the present phase convention, in which the diagonal elements of ${\cal V}_u$ and 
${\cal V}_d$ are chosen to be real. 
Also the other elements are given as
\begin{eqnarray}
  V_{cb}  & = &  x^{\a_2} \, N_2^{(u)} N_3^{(d)} \, 
             \left[ \frac{\bar{m}_{11}^* \, (\bar{\bz}_1^* \cdot {\bm}_3)}{m_{33} \, D_3^{11*}} \ 
                                               + \ x^{2(\a_1 - \a_2)} \times  \right.   \nonumber  \\
          &  &  \left.  \times \Bigl( f_{\a} \, n_{13}^* 
                 + \ \frac{| \bar{m}_{21} |^2 \, (\bar{\bz}_1^* \cdot {\bm}_3)}
                               {m_{33} \, \bar{m}_{11} \, D_3^{11*}}  \Bigr) + \O(\l^4) \right],    \\
  V_{ts}  & = &  - x^{\a_2} \, N_3^{(u)} N_2^{(d)} \, 
             \left[ \frac{\bar{m}_{11} \, (\bar{\bz}_1 \cdot {\bm}_3^*)}{m_{33}^* \, D_3^{11}} \ 
                                               + \ x^{2(\a_1 - \a_2)} \times  \right.   \nonumber  \\
          &  &  \left.  \times \Bigl( f_{\a}^* \, n_{13} 
                 + \ \frac{ \bar{m}_{21} \, l_{12} \, (\bar{\bz}_1 \cdot {\bm}_3^*)}
                               {m_{33}^* \, l_{11} \, D_3^{11}}  \Bigr) + \O(\l^4)  \right], 
\end{eqnarray}
where $x^{\a_2} \sim \l^2$. 
These equations yield the relation 
\begin{equation}
    \left| V_{ts} \right|^2 = \left| V_{cb} \right|^2 \times 
                           \left[ 1 - \left| V_{us} \right|^2 + p \, \l^4 \right] 
\label{eqn:Relation1}
\end{equation}
with $p = \O(1)$, which is free from the phase convention. 
From the unitarity condition on $V_{\rm CKM}$ \, Eq.(\ref{eqn:Relation1}) can be rewritten as 
\begin{equation}
    \left| V_{cd} \right|^2 = \left| V_{us} \right|^2 
                            \times \left[ 1 - \left| V_{cb} \right|^2 + \O(\l^6) \right]. 
\label{eqn:Relation2}
\end{equation}
Further we obtain 
\begin{eqnarray}
  V_{td}  & = &   x^{\a_1} \, N_3^{(u)} N_1^{(d)} \, 
          \left[ \frac{\xi_d^2 \, \bar{z}_{11}^* \, ({\bm}_3^* \cdot \bar{\bz}_1)}{{m}_{33}^* \, l_{11}} \ 
          + \ x^{2(\a_1 - \a_2)} \, \frac{\xi_d^2}{{m}_{33}^* \, (l_{11})^2} \times  \right.   \nonumber   \\
       &  &  \qquad  \times \Bigl( \xi_d^2 \, \bar{z}_{11}^* \, \bar{z}_{12}^* \, ({\bm}_3 \times {\bz}_3)_1^* + 
                      \ \bar{z}_{12}^* \, |\bar{m}_{11}|^2 \, ({\bm}_3^* \cdot \bar{\bz}_2) -    \nonumber  \\
       &  &  \qquad \qquad \qquad  \left.  - \ \frac{l_{21}}{l_{11}} \, \bar{m}_{11} \, 
                        D_3^{11*} \, ({\bm}_3^* \cdot \bar{\bz}_1) \Bigr)  + \O(\l^4) \right], \\  
  V_{ub}  & \simeq &  x^{3\a_1 - 2\a_2} \, N_1^{(u)} N_3^{(d)} \,
                         \frac{\xi_d^2 \, z_{33}^* \, ({\bz}_3 \cdot \bar{\bm}_1^*)}
                             {\bar{m}_{11}^* \, | D_3^{11}|^2}, 
\label{eqn:Vub}
\end{eqnarray}
where $x^{\a_1} \sim \l^3$ and $x^{3\a_1 - 2\a_2} \sim \l^5$. 
In the above equation (\ref{eqn:Vub}) the element $V_{ub}$ is rather small compared 
to the element $V_{td}$ due to the cancellation of leading term. 
The above expression of $V_{td}$ leads us to the relation 
\begin{equation}
    \left| V_{td} \right|^2 = \left| V_{us} \, V_{cb} \right|^2 \times ( 1+ q \, \l^2) 
\label{eqn:Relation3}
\end{equation}
with $q = \O(1)$.

\vspace{5mm}

\section{The Quark \textit{CP}-phase}

Here we pay attention to Eqs. (\ref{eqn:Relation1}) and 
(\ref{eqn:Relation3}), which yield interesting relations between 
the \textit{CP}-phase and three mixing angles. 
It has been shown that in the present model the hierarchical magnitude of 
three mixing angles can be understood systematically\cite{Matsu3}. 
The recent values of the CKM matrix elements have been summarized by PDG\cite{PDG} as 
\begin{eqnarray}
  & |V_{ts}| = 0.0405  \, \pm 0.0012,  \qquad  &   |V_{cb}| = 0.0414  \, \pm 0.0012,   \nonumber  \\
  & |V_{us}| = 0.22536 \, \pm 0.00061, \qquad  &  |V_{cd}| = 0.22522 \, \pm 0.00061,   \\
  & |V_{td}| = 0.00886 \, \pm 0.00033.  \qquad  &  \nonumber 
\end{eqnarray}
As seen in Eq.\eqref{CKMPDG}, the values of $|V_{ts}| $ and $|V_{cb}| $ are mainly determined 
by $s_{23}$ due to the hierarchical structure of the mixing angles.  
So it is noted that the double-sign in them corresponds in the same order. 
Then, with the aid of these data the relations (\ref{eqn:Relation1}) and (\ref{eqn:Relation3}) 
lead to 
\beq
    p = 3.0 \, \pm 0.6,  \qquad \qquad   q = -1.9 \, \pm 0.4, 
\eeq
which are consistent with $p, q = \O(1)$. 
The relation (\ref{eqn:Relation2}) is also in good agreement with the data. 
These results are in support of the present analyses up to the next-to-leading approximation 
in the F-N scheme.

In the standard representation of PDG\cite{PDG} for $V_{\rm CKM}$ 
the relation(\ref{eqn:Relation1}) becomes 
\beqa
\left| c_{12} \, s_{23} + s_{12} \, c_{23} \, s_{13} \, e^{i \, \delta} \right|^2 = 
        ( s_{23} \, c_{13})^2 \times \left[ 1 -  (s_{12} \, c_{13})^2 + p \, \l^4 \right], 
\label{eqn:Relation100}
\eeqa
which is rewritten as 
\begin{eqnarray}
  \cos \delta & = &  
       \frac{s_{23} \, (c_{13})^2}{2 \, s_{12} \, c_{12} \, c_{23} \, s_{13}} \, p \, \l^4 
        - \frac{s_{13}}{2 \, s_{12} \, c_{12} \, s_{23} \, c_{23}} 
           \Bigl( (s_{12} \, c_{23})^2 + (s_{23})^2    \nonumber  \\
       &  & \qquad \qquad    - (s_{23} \, s_{12})^2 \left( 1 + (c_{13})^2 \right) \Bigr). 
\label{eqn:delta1}
\end{eqnarray}
It is worth noting that in Eq.~\eqref{eqn:Relation100} the leading terms in the l.h.s.
and the r.h.s. are canceled out and that Eq.~\eqref{eqn:delta1} represents
a relation between the non-leading terms.
When we take $p = 3.0 \pm 0.6$ and the experimental values\cite{PDG}
\[
   s_{12} = 0.22536 \, \pm 0.00061, \qquad  
   s_{23} = 0.0414  \, \pm 0.0012,  \qquad  
   s_{13} = 0.00355 \, \pm 0.00015, 
\]
the above equation (\ref{eqn:delta1}) results in $ \cos \delta =0.20 \, \pm 0.04$. 
It might be thought that we can simply  derive $\cos\delta$ from 
$|V_{ts}|=\left| c_{12} \, s_{23} + s_{12} \, c_{23} \, s_{13} \, e^{i \, \delta} \right|$.
However, it is impossible for us to get information about $\cos\delta$
because the magnitude of the experimental error of $|V_{ts}|$ is larger than 
the coefficient of $ e^{i \delta}$.

Further Eq.(\ref{eqn:Relation3}) is translated into 
\beqa
   \left| s_{12} \, s_{23} - c_{12} \, c_{23} \, s_{13} \, e^{i \, \delta} \right|^2 = 
        ( s_{12} \, s_{23} \, c_{13}^2)^2 \times \left(1 + q \, \l^2 \right), 
\eeqa
which contains 
\beq
   \cos \delta =  
       \frac{c_{12} \, c_{23} \, s_{13}}{2 \, s_{12} \, s_{23}} 
         - \frac{s_{12} \, s_{23}}{2 \, c_{12} \, c_{23} \, s_{13}} 
              \Bigl(  (c_{13})^4 \, q \, \l^2 - (s_{13})^2 (1 + (c_{13})^2) \Bigr). 
\label{eqn:delta2}
\eeq
Substituting the experimental values of the three mixing angles and $q = -1.9 \pm 0.4$ 
to Eq.(\ref{eqn:delta2}), 
we obtain $\cos \delta =0.32 \, \pm 0.03$. 
Note that the uncertainty of $\cos\delta$ obtained here is rather small compared with 
that ($\cos\delta=0.32 \pm 0.11$) determined directly from  
$|V_{td}|=\left| s_{12} \, s_{23} - c_{12} \, c_{23} \, s_{13} \, e^{i \, \delta} \right|$.

Within $2 \sigma$ there is no discrepancy in the above two values of $\cos \delta$. 
Consequently, we conclude with $\cos \delta =0.26 \, \pm 0.05$ and 
$\delta = (75 \pm 3)^{\circ}$ in this analysis. 
The current experimental data show that $\gamma = (68 \pm 8)^{\circ}$ \cite{PDG}, 
where $\gamma = {\rm arg}\left( - (V_{ud} \, V_{ub}^*) / ( V_{cd} \, V_{cb}^*) \right) 
\simeq \delta$. 
In addition, the recent avaraged value is 
$\gamma = (73$$ + 9 \atop - 10$$)^{\circ}$ \cite{Kop}. 
So our result is consistent with the data.

\vspace{5mm}

\section{Summary}

In the present model, the characteristic structure of the CKM matrix is attributed to 
the hierarchical effective Yukawa couplings due to the Froggatt-Nielsen mechanism and 
also to the state-mixings beyond the MSSM. 
The $D^c$-$g^c$ mixings as well as generation mixings take place 
in the down-type quark sector . 
On the other hand, in the up-type quark sector we have no such mixings. 
These differences cause the nontrivial structure in CKM matrix. 
Specifically, the down-type mass matrix is described in terms of $\M$ and $\Z$ 
matrices in contrast with the up-type mass matrix of $\M$ itself. 
As a result, all off-diagonal elements of the CKM matrix are expressed as the products 
of $\M$ and $\Z$ elements.

In the $D^c$-$g^c$ mixings, since $D^c$ and $g^c$ are both $SU(2)_L$-singlets, 
the disparity between the diagonalization matrices for up-type quarks and 
down-type quarks in $SU(2)_L$-doublets is rather small. 
Accordingly, $V_{\rm CKM}$ exhibits small mixing. 
In this study we have found interesting relations between the \textit{CP}-phase and 
three mixing angles without relying on a specific flavor symmetry. 
Taking the current data of three mixing angles, 
we estimate the quark \textit{CP}-phase at $\delta = (75 \pm 3)^{\circ}$. 
This result is in accord with the current data of $\delta$.

The relations between the \textit{CP}-phase and three mixing angles stem from the fact 
that the CKM matrix comprises only two matrices $\M$ and $\Z$. 
This is because all matter fields belong to either of $({\bf 15},{\bf 1})$ or 
$(\overline{\bf 6},{\bf 2})$ representations in the gauge group $SU(6) \times SU(2)_R$ . 
If the gauge group is chosen to be smaller than the above group, 
the number of irreducible representations for matter fields becomes larger than two. 
In such a case, there appear more parameters and hence we have no interesting relations 
between the \textit{CP}-phase and three mixing angles. 
We expect that the present model gives a comprehensive explanation of 
fermion mass spectra and mixing matrices together with the longevity of the proton 
and gauge coupling unification\cite{Matsu1,Matsu2,Matsu3,Matsu4,Matsu5}.

Finally, we touch upon the study of the the MNS matrix\cite{MNS}. 
The observed features of the MNS matrix differ considerably from those of 
the CKM matrix. 
In the present model the $L$-$H_d$ mixings occur in the lepton sector. 
Since $L$ and $H_d$ are both $SU(2)_L$-doublets, 
there appears no disparity between the diagonalization matrices for charged leptons 
and neutrinos unless the seesaw mechanism does not take place. 
As a matter of fact, however, the seesaw mechanism is at work and an additional 
transformation is required to diagonalize the neutrino mass matrix. 
This additional transformation matrix yields nontrivial $V_{\rm MNS}$. 
The seesaw mechanism brings about the cancellation of the F-N factors 
in the neutrino mass matrix\cite{Matsu5, HMM1}. 
As a consequence, there is no hierarchical structure in neutrino mass matrix 
and eventually $V_{\rm MNS}$ exhibits large mixing. 
It does not seem that the MNS matrix elements are connected to the CKM matrix elements 
in an uncomplicated way. 
For this reason it is difficult for us to find a simple interrelation 
among the quark \textit{CP}-phase and the leptonic \textit{CP}-phase.

\vspace{5mm}


\end{document}